\documentclass[12pt,showpacs]{revtex4} %Updated Sep 24, 2009%
\usepackage{amssymb,epsf}
\usepackage{latexsym}
\usepackage{color}
\begin{document}
\title{\textbf{Dirac quantization of noncommutative Abelian Proca field}}
\author{F. Darabi}
\email{f.darabi@azaruniv.edu} 
\author{F. Naderi}
\affiliation{Department of Physics, Azarbaijan University of Tarbiat Moallem, Tabriz, Iran, P.O. Box: 53714-161,}

\affiliation{Research Institute for Astronomy and Astrophysics of
Maragha (RIAAM)- Maragha, Iran, P.O. Box: 55134-441.}
\date{\today}
\begin{abstract}
\textbf{Abstract:} 
Dirac formalism of Hamiltonian constraint systems is studied for the noncommutative Abelian Proca field. It is shown that the system of constraints are of second class in agreement with the fact that the Proca field is not guage invariant. Then, the system of second class constraints is quantized by introducing Dirac brackets in the reduced phase space.


\end{abstract}

\maketitle
\newpage
\section{Introduction}

The noncommutative geometry, pioneered by alain Connes, aims at
a generalization of geometrical ideas to spaces whose coordinates fail to
commute \cite{Con}. In this theory it is postulated that space-time is noncommutative at very high energies and one tries to guess the small scale structure of space-time from our present knowledge at the electroweak scale. String theory on the other hand aims at deriving the standard model directly from the Planck scale physics. Thus, noncommutative geometry may describe the low energy dynamics in string theory as a picture for the standard model, where symmetries act directly by a group of coordinate transformations on an underlying space-time manifold producing the electroweak and strong forces as pseudo-forces. Moreover, it may place gravity and the other forces on the same footing by obtaining all forces as pseudo-forces from some general coordinate transformations acting on some general spacetime.

Quantum field theories on noncommutative spaces are usually formulated in terms of star products of ordinary functions \cite{Moy}. The different field models on noncommutative spaces has been recently of particular interest due to the recent development of the superstring theory. It was shown that
noncommutative coordinates emerge naturally in the perturbative version of the D-brane theory, namely low energy excitations of a D-brane, with the presence of the external background magnetic field \cite{Sei}.
Therefore, noncommutative Yang-Mills theories appear in the string theory
in an effective way and that is why they are being so widely investigated.
It is known that there exists a Seiberg-Witten map between noncommutative field theories and effective commutative theories in that both have the same degrees of freedom. Moreover, noncommutative gauge theories can be represented as ordinary gauge theories with the same degrees of freedom, and with the
additional deformation parameter $\theta$ \cite{Mad}. 
The Seiberg-Witten map between field theory on noncommutative spaces and the corresponding commutative field theory allows one to formulate an action
principle in terms of ordinary field. The effective Lagrangian of this action is expanded as series of ordinary field and the noncommutative parameter $\theta$ which plays the role of coupling constant.

Attempts to quantize the Maxwell theory on noncommutative spaces have already
been done. First, the corresponding commutative action in terms of ordinary fields and linear in the deformation parameter has been derived in \cite{Jur, Bic}. Afterwards, the Dirac's quantization of Hamiltonian constraint systems \cite{Dir} has been applied on this commutative action in \cite{Kru}.
,  
In the present paper, motivated by the quantization of Maxwell theory on noncommutative space, we attempt to quantize the noncommutative massive Abelian Maxwell theory, namely noncommutative Abelian Proca field, using the Dirac quantization procedure in a similar way as in Ref.\cite{Kru}. In section II, we first obtain the noncommutative action for the Proca field using the Moyal product and then derive the corresponding commutative action in terms of ordinary fields and linear in the deformation parameter $\theta$. In section III, we study this action in the context of the Dirac's Hamiltonian constraint systems to find the corresponding constraints which turn out to
be the second class type. In section IV, the system of second class constraints are quantized in the reduced phase space. The paper ends with a conclusion.   

\section{Noncommutative Abelian Proca field}

The action of Abelian noncommutative Proca field is written 
\begin{equation}\label{1}
S=\int(-\frac{1}{4}\hat{F}_{\mu \nu}\ast \hat{F}_{\mu \nu}+\frac{1}{2}m^2\hat{A}_{\mu}\ast
\hat{A}_{\nu})d^4x,
\end{equation}
where $\ast$ denotes the {\it Star product}, $\hat{A}_{\mu}$ and $\hat{F}_{\mu \nu}$ are the vector potential and field strength tensor respectively, and $m$ is the mass of the $\hat{A}_{\mu}$ field. The fields $\hat{F}_{\mu \nu}$ and $\hat{A}_{\mu}$ may be expressed in terms of the corresponding commutative
quantities as follows \cite{Bic}
\begin{equation}\label{2}
\hat{A}_{\mu}={A}_{\mu}-\frac{1}{2}\theta_{\alpha \beta}A_{\alpha}(\partial
_{\beta}{A}_{\mu}+F_{\beta \mu})
\end{equation}
\begin{equation}\label{3}
\hat{F}_{\mu \nu}=\partial_{\mu}\hat{A}_{\nu}-\partial_{\nu}\hat{A}_{\mu}-i
\hat{A}_{\mu}\ast \hat{A}_{\nu}+i
\hat{A}_{\nu}\ast \hat{A}_{\mu},
\end{equation}
where $\theta_{\alpha \beta}$ stands for the noncommutativity parameter that characterizes non-commutativity through the coordinate commutation relation $[x_{\alpha}, x_{\beta}]=i\theta_{\alpha \beta}$ \cite{Con}, \cite{Syn}. It is known
that the integral over the star product of quantities is equal to the corresponding
integral over the ordinary product \cite{Ria}, then we may rewrite the action (\ref{1})
as
\begin{equation}\label{4}
S=\int(-\frac{1}{4}\hat{F}_{\mu \nu} \hat{F}_{\mu \nu}+\frac{1}{2}m^2\hat{A}_{\mu}
\hat{A}_{\nu})d^4x.
\end{equation}
Using (\ref{2}) and (\ref{3}) the Lagrangian theory of the above noncommutative
action may be expanded as the commutative theory with the same degrees of freedom, and with the additional terms containing the noncommutative parameter
$\theta_{\alpha \beta}$ of the first order
\begin{eqnarray}\label{5}
\hat{{\cal L}}&=&-\frac{1}{4}{F}_{\mu \nu} {F}_{\mu \nu}+\frac{1}{8}\theta_{\alpha \beta}F_{\alpha \beta}{F}_{\mu \nu}^2-\frac{1}{2}\theta_{\alpha \beta}F_{\mu
\alpha}{F}_{\nu \beta}{F}_{\mu \nu}\\ \nonumber
&+&\frac{1}{2}m^2(A_{\mu}^2-\theta_{\alpha \beta}A_{\alpha}(\partial
_{\beta}{A}_{\mu}+F_{\beta \mu})A_{\mu}),
\end{eqnarray} 
where the commutative field strength tensor is
\begin{equation}\label{6}
{F}_{\mu \nu}=\partial_{\mu}{A}_{\nu}-\partial_{\nu}{A}_{\mu}.
\end{equation}
Now we define the followings
\begin{equation} \label{7}
\left\{ \begin{array}{ll} A_{\mu}=(\vec{A}, i A_0)
\\
E_i=iF_{i4}
\\
B_i=\frac{1}{2}\epsilon_{ijk}F_{jk}
\\
\theta_i=\frac{1}{2}\epsilon_{ijk}\theta_{jk}.
\end{array}
\right.
\end{equation}
Then, the Lagrangian density (\ref{5}) casts in the following form 
\begin{eqnarray}\label{8}
\hat{{\cal L}}&=&\frac{1}{2}(E^2-B^2)(1+\vec{\theta} \cdot \vec{B})-(\vec{\theta} \cdot \vec{E})(\vec{E}
\cdot \vec{B})+\frac{m^2}{2}(-A_0^2+A_i^2)
\\ \nonumber
&+&\frac{m^2}{4}(\vec{\theta} \times \vec{A})\cdot \vec{\nabla}(A_0^2)-\frac{m^2}{2}[(\vec{\theta} \times \vec{A})\cdot \vec{E}]A_0+3\frac{m^2}{4}[(\vec{\theta} \cdot \vec{B})A_j^2-(\vec{\theta} \cdot \vec{A})(\vec{A} \cdot \vec{B})].
\end{eqnarray} 
The Euler-Lagrange equations for the noncummutative Proca field are obtained
\begin{eqnarray}\label{9}
\partial_{\rho}F_{\rho \sigma}&-&\frac{1}{4}\partial_{\rho}(\theta_{\rho \sigma}F_{\mu
\nu}^2)-\frac{1}{2}\partial_{\rho}(\theta_{\alpha \beta}F_{\alpha
\beta}F_{\rho \sigma})+\partial_{\rho}(\theta_{\sigma \beta}F_{\nu
\beta}F_{\rho \nu})\\ \nonumber
&-&\partial_{\rho}(\theta_{\rho \beta}F_{\nu
\beta}F_{\sigma \nu})+\frac{1}{2}\partial_{\rho}(\theta_{\alpha \beta}F_{\rho
\alpha}F_{\sigma \beta})-\frac{1}{2}\partial_{\rho}(\theta_{\alpha \beta}F_{\sigma
\alpha}F_{\rho \beta})\\ \nonumber
&-&\frac{m^2}{2}\theta_{\sigma \beta}(\partial
_{\beta}{A}_{\mu}+F_{\beta \mu})A_{\mu}-\frac{m^2}{2}\theta_{\alpha \beta}(\partial
_{\beta}{A}_{\sigma}+F_{\beta \sigma})A_{\alpha}\\ \nonumber
&+&m^2\partial_{\rho}(\theta_{\alpha \rho}A_{\alpha}A_{\sigma})-\frac{m^2}{2}\partial_{\rho}(\theta_{\alpha \sigma}A_{\alpha})A_{\rho}+m^2A_{\sigma}=0,
\end{eqnarray}  
where in the last line the Lorentz condition $\partial_{\mu}A_{\mu}=0$ has been used using the fact that this condition holds in the massive as well as massless Maxwell field \cite{Ryder}.
Using (\ref{7}), the field equations are divided into the following two set
of equations 
\begin{equation}\label{10}
\frac{\partial \vec{D}}{\partial t}-\vec{\nabla} \times\vec{H}=-\vec{J},
\end{equation}
\begin{equation}\label{11}
\vec{\nabla} \cdot \vec{D}=\rho,
\end{equation}
where $(\vec{\nabla} \times\vec{H})_i=\epsilon_{ijk}\partial_jH_k$, $\vec{\nabla} \cdot
\vec{D}=\partial_i
D_i$, $\partial/\partial t=i\partial_4$ and 
\begin{equation}\label{12}
\rho=m^2[A_0+\frac{1}{2}\vec{\nabla} \cdot (\vec{\theta} \times \vec{A})A_0+\frac{1}{2} (\vec{\theta} \times \vec{A})\cdot \vec{E}],
\end{equation}
\begin{equation}\label{13}
\vec{D}=\vec{E}+(\vec{\theta} \cdot \vec{B})\vec{E}-(\vec{\theta} \cdot \vec{E})\vec{B}-(\vec{E} \cdot \vec{B})\vec{\theta}-\frac{m^2}{2}(\vec{\theta} \times \vec{A})A_0,
\end{equation}
\begin{equation}\label{14}
\vec{J}=m^2[\vec{A}-\frac{1}{2}(\vec{E} \times \vec{\theta})A_0+\frac{3}{2}(\vec{\theta} \cdot \vec{B})\vec{A}-\frac{3}{4}(\vec{\theta} \cdot \vec{A})\vec{B}-\frac{3}{4}(\vec{A} \cdot \vec{B})\vec{\theta}],
\end{equation}
\begin{equation}\label{15}
\vec{H}=\vec{B}+(\vec{\theta} \cdot \vec{B})\vec{B}+(\vec{\theta} \cdot \vec{E})\vec{E}-
\frac{1}{2}(E^2-B^2)\vec{\theta}-m^2(\frac{1}{4}A_0^2+A_j^2)\vec{\theta}+m^2\vec{A}(\vec{\theta} \cdot \vec{A}).
\end{equation}
On the other hand, using the strength tensor $F_{\mu \nu}=\partial_{\mu}A_{\nu}-\partial_{\nu}A_{\mu}$ the source-less equations are 
\begin{equation}\label{16}
\partial_{\mu}\tilde{F}_{\mu \nu}=0,
\end{equation}
where $\tilde{F}_{\mu \nu}=\frac{1}{2}\epsilon_{\mu \nu \alpha \beta}F_{\alpha \beta}$ and $\epsilon_{\mu \nu \alpha \beta}$ is the Levi-Civita constant
tensor ($\epsilon_{1234}=-i$). 
Therefore, equations (\ref{16}) may be written as 
\begin{equation}\label{17}
\frac{\partial \vec{B}}{\partial t}+\vec{\nabla} \times\vec{E}=0,
\end{equation}
\begin{equation}\label{18}
\vec{\nabla} \cdot \vec{B}=0.
\end{equation}

\section{Hamiltonian constraint system approach}

In this section, following Dirac, we will study the dynamics of the
noncommutative Abelian Proca field in the context of Hamiltonian constraint systems \cite{Dir}.
In so doing, we first obtain the conjugate momenta of $A_i$ and $A_0$, respectively
as 
\begin{equation}\label{19}
\pi_i=\frac{\partial\hat{{\cal L}}}{\partial(\partial_0A_i)}=-E_i(1+\vec{\theta} \cdot \vec{B})+(\vec{\theta} \cdot \vec{E})B_i+(\vec{E}
\cdot \vec{B})\theta_i+\frac{m^2}{2}(\vec{\theta} \times \vec{A})_i A_0,
\end{equation}
\begin{equation}\label{20}
\pi_0=\frac{\partial\hat{{\cal L}}}{\partial(\partial_0A_0)}=0.
\end{equation}
Equation (\ref{20}) results in a primary constraint
\begin{equation}\label{21}
\phi_1\approx 0,
\end{equation}
where $\phi_1\equiv \pi_0$, and comparison of (\ref{13}) with (\ref{19})
leads to  
\begin{equation}\label{22}
\pi_i-=D_i.
\end{equation}
Therefore, we obtain the following commutation relations
\begin{equation}\label{23}
\{A_i(x,t), D_j(y,t)\}=-\delta_{ij}\delta(x-y),
\end{equation}
\begin{equation}\label{24}
\{B_i(x,t), D_j(y,t)\}=\epsilon_{ijk}\partial_k\delta(x-y).
\end{equation}
Using the Legendre transformation as ${\cal H}=\pi_{\mu}\partial_0 A_{\mu}-{\cal L}$, we may obtain the Hamiltonian density
\begin{eqnarray}\label{25}
\hat{{\cal H}}_0&=&\frac{1}{2}(E^2+B^2)(1+\vec{\theta} \cdot \vec{B})-(\vec{\theta} \cdot \vec{E})(\vec{E}
\cdot \vec{B})+\frac{m^2}{2}A_0^2-\frac{m^2}{2}A_i^2(1+\frac{3}{2}\vec{\theta} \cdot \vec{B})
\\ \nonumber
&-&\frac{m^2}{2}(\vec{\theta} \times \vec{A})_i (\partial_i A_0)A_0+3\frac{m^2}{4}(\vec{\theta} \cdot \vec{A})(\vec{A} \cdot \vec{B})-\pi_i \partial_i A_0.
\end{eqnarray} 
Using (\ref{19}), we may obtain $E_i$ in terms of $\pi_i$ to first order
in $\theta$ as
\begin{equation}\label{26}
E_i=-\pi_i(1-\vec{\theta} \cdot \vec{B})-(\vec{\pi} \cdot \vec{B})B_i-(\vec{\pi}
\cdot \vec{B})\theta_i+\frac{m^2}{2}(\vec{\theta} \times \vec{A})_i A_0.
\end{equation}
Substituting for $E_i$ in Eq.(\ref{25}) in terms of $\pi_i$ we obtain
\begin{eqnarray}\label{27}
\hat{{\cal H}}_0&=&\frac{1}{2}(\pi^2+B^2)+\frac{1}{2}(B^2-\pi^2)(\vec{\theta} \cdot \vec{B})+(\vec{\pi} \cdot \vec{\theta})(\vec{B} \cdot \vec{\pi})
+\frac{m^2}{2}A_0^2-\frac{m^2}{2}(\vec{\theta} \times \vec{A})\cdot \vec{\pi} A_0
\\ \nonumber
&-&\frac{m^2}{2}A_i^2(1+\frac{3}{2}\vec{\theta} \cdot \vec{B})-\frac{m^2}{2}(\vec{\theta} \times \vec{A})_i (\partial_i A_0)A_0+3\frac{m^2}{4}(\vec{\theta} \cdot \vec{A})(\vec{A} \cdot \vec{B})-\pi_i \partial_i A_0 +{\cal O}(\theta^2).
\end{eqnarray}  
Now, we study the consistency condition for the primary constraint (\ref{21})
\begin{equation}\label{28}
\dot{\phi}_1=\{\phi_1(x), H_0\}= 0,
\end{equation}
where 
\begin{equation}\label{29}
H_0=\int \hat{{\cal H}}_0(x)d^3x.
\end{equation}
The consistency condition (\ref{28}) results in a secondary constraint 
\begin{equation}\label{30}
\phi_2=\partial_i \pi_i+m^2 A_0+\frac{m^2}{2}\vec{\nabla} \cdot (\vec{\theta} \times \vec{A})A_0-\frac{m^2}{2} (\vec{\theta} \times \vec{A})\cdot \vec{\pi},
\end{equation}
or
\begin{equation}\label{31}
\phi_2=\partial_i \pi_i+m^2 A_0+\frac{m^2}{2}\vec{\nabla} \cdot (\vec{\theta} \times \vec{A})A_0+\frac{m^2}{2} (\vec{\theta} \times \vec{A})\cdot \vec{E}+{\cal O}(\theta^2).
\end{equation}
If we put $\pi_i=-D_i$, this equation casts in the form of generalized Gauss law which has already been obtained in (\ref{11}).
Now, the extended Hamiltonian is constructed by adding the primary constraint
$\phi_1$ up to an arbitrary coefficient $u(x)$
\begin{equation}\label{34}
H_T=H_0+\int \, u(x) \phi_1(x) d^3x.
\end{equation}
The consistency condition for the secondary constraint $\phi_2$ is now considered
as
\begin{equation}\label{35}
\dot{\phi}_2=\{\phi_2(x), H_T\}= 0,
\end{equation}
which introduces no new constraint and just fixes the unknown coefficient as
\begin{eqnarray}\label{36}
u(x)&=&-\vec{\nabla} \cdot (\vec{\theta} \times \vec{A})A_0-\partial_i\left(A_i\left(
1+\frac{3}{2}\,\vec{\theta} \cdot \vec{B}-\frac{1}{2}\vec{\nabla} \cdot (\vec{\theta} \times \vec{A})\right) \right)\\ \nonumber
&+&\frac{3}{4}\,\partial_i(\theta_i(\vec{A} \cdot \vec{B})+B_i(\vec{\theta} \cdot \vec{A}) )+(\vec{\theta} \times \vec{A})\cdot
(\vec{\nabla} \times \vec{B})+{\cal O}(\theta^2),
\end{eqnarray}
where we have used the expansion $\frac{1}{1+\epsilon}\simeq (1-\epsilon)$ due to
the smallness of $\vec{\theta}$. It is easy to show that the constraints $\phi_1$ and $\phi_2$ are second class, namely
\begin{eqnarray}\label{32}
\{\phi_1(x), \phi_2(x') \}=-m^2(1+\frac{1}{2}\vec{\nabla} \cdot (\vec{\theta} \times \vec{A}))\delta(x-x')
\end{eqnarray}
or
\begin{eqnarray}\label{33}
\{\phi_i(x), \phi_j(x') \}=\left(\begin{array}{cc} 0 & -m^2(1+\frac{1}{2}\vec{\nabla} \cdot (\vec{\theta} \times \vec{A}))\\
\\m^2(1+\frac{1}{2}\vec{\nabla} \cdot (\vec{\theta} \times \vec{A})) & 0 \end{array}\right)\delta(x-x').
\end{eqnarray}
It is of great importance to distinction between first and second class constraints. The first class constraints are defined as the constraints which "commute" (i.e. have vanishing Poisson brackets) with all the other constraints. This situation brings to light the presence of some gauge degrees of freedom in the Dirac formalism. On the other hand, the second class constraints have at least one non vanishing bracket with some other constraints, like (\ref{32}).\\
By manipulation and integration by parts
in some appropriate terms in (\ref{27}) the constraint
$\phi_2$ appears in the Hamiltonian as
\begin{eqnarray}\label{37}
H_0&=&\int \, [\frac{1}{2}(\pi^2+B^2)+\frac{1}{2}(B^2-\pi^2)(\vec{\theta} \cdot \vec{B})+(\vec{\pi} \cdot \vec{\theta})(\vec{B} \cdot \vec{\pi})
-\frac{m^2}{2}A_0^2
\\ \nonumber
&-&\frac{m^2}{2}A_i^2(1+\frac{3}{2}\vec{\theta} \cdot \vec{B})
+\frac{m^2}{4}\vec{\nabla} \cdot (\vec{\theta} \times \vec{A})A_0^2+
3\frac{m^2}{4}(\vec{\theta} \cdot \vec{A})(\vec{A} \cdot \vec{B})+A_0 \phi_2]d^3x.
\end{eqnarray}  
Since the original Hamiltonian $H_0$ includes the constraint $\phi_2$ with the known coefficient $A_0$, it is not necessary to add once again this constraint with an unknown coefficient to the Hamiltonian $H_T$, and so we obtain the extended
Hamiltonian  
\begin{equation}\label{38}
H_E=\bar{H}_0+\int \, [u(x) \phi_1(x) +  A_0(x) \phi_2(x)] d^3x,
\end{equation}
where
\begin{eqnarray}\label{39}
\bar{H}_0&=&\int \, [\frac{1}{2}(\pi^2+B^2)+\frac{1}{2}(B^2-\pi^2)(\vec{\theta} \cdot \vec{B})+(\vec{\pi} \cdot \vec{\theta})(\vec{B} \cdot \vec{\pi})
-\frac{m^2}{2}A_0^2
\\ \nonumber
&-&\frac{m^2}{2}A_i^2(1+\frac{3}{2}\vec{\theta} \cdot \vec{B})
+\frac{m^2}{4}\vec{\nabla} \cdot (\vec{\theta} \times \vec{A})A_0^2+
3\frac{m^2}{4}(\vec{\theta} \cdot \vec{A})(\vec{A} \cdot \vec{B})]d^3x.
\end{eqnarray} 

\section{Quantization of second class constraint system}

The problem of quantization of Hamiltonian constraint systems is twofold:
quantization of first class constraints and of second class
constraints \cite{Dir}. However, the problem of quantizing theories with second class constraints is less ambiguous than quantizing theories with first class constraints. In the following analysis we will not deal with first class constraints due to the fact that the Proca field is not gauge invariant. In the case of second class constraints we can switch to new canonical brackets in order to set all of the second class constraints strongly equal to zero. This means that in any given quantity, such as the Hamiltonian, we can set them to zero "by hand". In such a case, we can safely change to the new canonical brackets, the so called Dirac brackets, defined as follows
\begin{eqnarray}\label{40}
\{A(x, t), B(y, t)\}_{DB}&=&\{A(x, t), B(y, t)\}
\\ \nonumber
&-&\int \,\{A(x, t), \phi_{i}(z, t)\}C_{ij}^{-1}(z, \omega)\{\phi_{j}(\omega, t), B(y, t)\}\,d^3z\,d^3 \omega,
\end{eqnarray}
where $C_{ij}(x, z)=\{\phi_{i}(x), \phi_{j}(z)\}$ is given by (\ref{33}). We may obtain the inverse matrix $C^{-1}_{ij}$ using
the relation
\begin{equation}\label{41}
\int\,C_{ij}(x, z)C^{-1}_{jk}(z, y)\,d^3z=\delta_{ik}\delta(x-y), 
\end{equation}
which leads to
\begin{eqnarray}\label{42}
C^{-1}(z, y)=\left(\begin{array}{cc} 0 & m^{-2}(1+\frac{1}{2}\vec{\nabla} \cdot (\vec{\theta} \times \vec{A}))^{-1}\\
\\ -m^{-2}(1+\frac{1}{2}\vec{\nabla} \cdot (\vec{\theta} \times \vec{A}))^{-1} & 0 \end{array}\right)\delta(z-y).
\end{eqnarray}
Therefore, the following Dirac brackets are obtained
\begin{equation}\label{43}
\{\pi_0(x, t), A_0(y, t)\}_{DB}=0, 
\end{equation}
\begin{equation}\label{44}
\{\pi_0(x, t), A_i(y, t)\}_{DB}=0, 
\end{equation}
\begin{equation}\label{45}
\{\pi_{\mu}(x, t), \pi_{\nu}(y, t)\}_{DB}=0, 
\end{equation}
\begin{equation}\label{46}
\{A_0(x, t), A_0(y, t)\}_{DB}=0, 
\end{equation}
\begin{equation}\label{47}
\{\pi_i(x, t), A_j(y, t)\}_{DB}=-\delta_{ij}\delta(x-y), 
\end{equation}
\begin{equation}\label{48}
\{\pi_i(x, t), B_j(y, t)\}_{DB}=-\epsilon_{ijk}\partial_k\delta(x-y), 
\end{equation}
\begin{eqnarray}\label{49}
\{A_0(x, t), A_j(y, t)\}_{DB}
&=&m^{-2}(1-\frac{1}{2}\vec{\nabla} \cdot (\vec{\theta} \times \vec{A}))\partial_j(x)\delta(x-y)\\ \nonumber
&-&\frac{1}{2} (\vec{\theta} \times \vec{A}))_j(x)\delta(x-y)+{\cal O}(\theta^2), \end{eqnarray}
\begin{eqnarray}\label{50}
\{A_0(x, t), B_k(y, t)\}_{DB}
&=&\epsilon_{klj}\partial_l(y)[m^{-2}(1-\frac{1}{2}\vec{\nabla} \cdot (\vec{\theta} \times \vec{A}))\partial_j(x)\delta(x-y)\\ \nonumber
&-&\frac{1}{2} (\vec{\theta} \times \vec{A}))_j(x)\delta(x-y)], 
\end{eqnarray}
\begin{eqnarray}\label{51}
\{\pi_i(x, t), A_0(y, t)\}_{DB}
&=&\frac{1}{2}[A_0(y)\epsilon_{jli}\theta_l(y)\partial_j(y)\delta(x-y)\\ \nonumber
&-&(\vec{\pi}\times
\vec{\theta})_i(y)\delta(x-y)]+{\cal O}(\theta^2). 
\end{eqnarray}
It is known that the Dirac brackets of second class constraints with each
arbitrary function $f$ is strongly zero, namely $\{\phi_i(x, t), f(y, t)\}_{DB}=0$.
Therefore, the second class constraints are supposed to be strongly zero
\cite{Dir} and so we obtain the physical Hamiltonian in the reduced phase space as
\begin{eqnarray}\label{52}
H_{Ph}=\bar{H}_0&=&\int \, [\frac{1}{2}(\pi^2+B^2)+\frac{1}{2}(B^2-\pi^2)(\vec{\theta} \cdot \vec{B})+(\vec{\pi} \cdot \vec{\theta})(\vec{B} \cdot \vec{\pi})
-\frac{m^2}{2}A_0^2
\\ \nonumber
&-&\frac{m^2}{2}A_i^2(1+\frac{3}{2}\vec{\theta} \cdot \vec{B})
+\frac{m^2}{4}\vec{\nabla} \cdot (\vec{\theta} \times \vec{A})A_0^2+
3\frac{m^2}{4}(\vec{\theta} \cdot \vec{A})(\vec{A} \cdot \vec{B})]d^3x.
\end{eqnarray} 
The equations of motion are obtained 
\begin{eqnarray}\label{53}
\dot{A}_i&=&\partial_0A_i=\{A_i, H_{Ph}\}_{DB}\\ \nonumber
&=&\pi_i(1-\vec{\theta} \cdot \vec{B})+
\theta_i(\vec{B} \cdot \vec{\pi})+(\vec{\pi} \cdot \vec{\theta})B_i-\partial_i(A_0-\vec{\nabla} \cdot (\vec{\theta} \times \vec{A}))-\frac{m^2}{2}A_0(\vec{\theta} \times \vec{A})_i, 
\end{eqnarray}
\begin{eqnarray}\label{54}
\dot{A}_0&=&\partial_0A_0=\{A_0, H_{Ph}\}_{DB}\\ \nonumber
&=&\frac{A_0}{2}\vec{\nabla} \cdot (\vec{\pi} \times \vec{\theta})
-\frac{1}{2}(\vec{\theta} \times \vec{A})\cdot (\vec{\nabla} \times\vec{B})-\partial_j(A_j(1+\frac{3}{2}\vec{\theta} \cdot \vec{B}))\\ \nonumber
&+&\frac{1}{2}\vec{\nabla} \cdot (\vec{\theta} \times \vec{A})\partial_j
A_j+\frac{3}{4}\partial_j(\theta_j(\vec{A} \cdot \vec{B})+(\vec{\theta} \cdot \vec{A})B_j)+{\cal O}(\theta^2).
\end{eqnarray}
\begin{eqnarray}\label{55}
\dot{\pi}_i&=&\partial_0\pi_i=\{\pi_i, H_{Ph}\}_{DB}\\ \nonumber
&=&-\vec{\nabla} \times[\vec{B}(1+\vec{\theta} \cdot \vec{B})+\frac{1}{2}\vec{\theta}(B^2-\pi^2)+\vec{\pi}(\vec{\pi}
\cdot \vec{\theta})-\frac{3}{4}m^2A_j^2\vec{\theta}-\frac{1}{4}m^2A_0\vec{\theta}-\frac{3}{4}m^2(\vec{\theta} \cdot \vec{A})A]_i\\ \nonumber
&+&m^2 A_i(1+\frac{3}{2}(\vec{\theta} \cdot \vec{B}))+\frac{1}{2}m^2A_0^2(\vec{\pi}
\times \vec{\theta})_i-\frac{3}{4}m^2(\theta_i(\vec{A} \cdot \vec{B})+(\vec{\theta} \cdot \vec{A})B_i).
\end{eqnarray}
Substituting for $\pi_i$ in (\ref{55}) in terms of $E_i$ through (\ref{19}),
ignoring the terms of the order ${\cal O}(\theta^2)$, using (\ref{14}), (\ref{15}) and
considering $\pi_i=-D_i$ we obtain
\begin{equation}\label{56}
\frac{\partial \vec{D}}{\partial t}-\vec{\nabla} \times\vec{H}=-\vec{J},
\end{equation}
which is the Ampere's Law. This means the equation of motion for $\pi_i$
in noncommutative theory obtained by Dirac formalism is in agreement with the Maxwell's equation.
On the other hand, equation (\ref{54}) may be rewritten as 
\begin{eqnarray}\label{57}
\partial_j(\frac{A_0}{2}(\vec{\pi} \times \vec{\theta})_j-A_j-\frac{3}{2}A_j(\vec{\theta}\cdot \vec{B})&+&\frac{3}{4}(\theta_j(\vec{A} \cdot \vec{B})+(\vec{\theta} \cdot \vec{A})B_j))\\ \nonumber
&=&\dot{A}_0+\frac{1}{2}(\vec{\theta} \times \vec{A})\cdot (\vec{\nabla} \times\vec{B})-\frac{1}{2}\vec{\nabla} \cdot (\vec{\theta} \times \vec{A})\partial_j
A_j.
\end{eqnarray}
The left hand side of above equation, using (\ref{14}), is equal to $-m^{-2}\vec{\nabla} \cdot \vec{J}$. The time derivative of $\rho$ in (\ref{12}) leads us to
\begin{equation}\label{58}
\dot{\rho}=m^2[\dot{A}_0+\frac{1}{2}\partial_j(\vec{\theta} \times \vec{\dot{A}})_j
A_0+\frac{1}{2}\partial_j(\vec{\theta} \times \vec{A})_j \vec{\dot{A}}_0+
\frac{1}{2} (\vec{\theta} \times \vec{A})\cdot \vec{\dot{E}}+\frac{1}{2} (\vec{\theta} \times \vec{\dot{A}})\cdot \vec{E}],
\end{equation}
which becomes the following form using (\ref{26}) and ignoring terms of
the order ${\cal O}(\theta^2)$
\begin{equation}\label{59}
\dot{\rho}=m^2[\dot{A}_0+\frac{1}{2}\partial_j(\vec{\theta} \times \vec{\dot{A}})_j
A_0+\frac{1}{2}\partial_j(\vec{\theta} \times \vec{A})_j \vec{\dot{A}}_0-
\frac{1}{2} (\vec{\theta} \times \vec{A})\cdot \vec{\dot{\pi}}-\frac{1}{2} (\vec{\theta} \times \vec{\dot{A}})\cdot \vec{\pi}].
\end{equation}
Finally, using the equations of motion (\ref{53}) and (\ref{55}) we obtain
\begin{equation}\label{60}
\dot{\rho}=m^2\left(\dot{A}_0+\frac{1}{2}(\vec{\theta} \times \vec{A})\cdot (\vec{\nabla} \times\vec{B})-\frac{1}{2}\vec{\nabla} \cdot (\vec{\theta} \times \vec{A})\partial_j
A_j \right).
\end{equation}
Therefore equation (\ref{57}) casts in the following form
\begin{equation}\label{61}
\frac{\partial{\rho}}{\partial t}+\vec{\nabla} \cdot \vec{J}=0.
\end{equation}
This shows that the equation of motion for $A_0$ in noncommutative
theory obtained by Dirac formalism is in complete agreement with the charge conservation. The quantization of system is then achieved by the standard replacement in the classical equations (\ref{53}), (\ref{54}) and (\ref{55}), namely
\begin{equation}\label{62}
\{\,,\}_{DB} \rightarrow \frac{1}{i\hbar} \,[\,, ]_{QM}.
\end{equation}  
\newpage
\section{Conclusion}

The systems with constraints are usually described by the Dirac's elegant formulation of Hamiltonian constraint systems. Two kinds of first and second
class constraints play the main role in this formulation. The first class
constraints are responsible for gauge invariance whereas the second class
constraints provides us with a reduced phase space. Linear combinations of second class constraints may result in the first class constraints. From
classical point of view no preference is made between these two kinds of constraints. However, the problem of quantizing theories with second class constraints is less ambiguous than quantizing theories with first class constraints.
We have studied the problem of quantization of the noncommutative Abelian Proca field. The noncomutative Abelian Proca field has been rewritten as a commutative theory containing the noncommutative tensor $\theta_{\alpha \beta}$ as a given external tensor (this is a breaking of manifest covariance) at the first order. Since the theory is massive, gauge invariance is broken
and instead of two first class constraints like in Maxwell theory (so
that $A_0$ and one component of $A_i$ are gauge variables) there is a pair
of second class constraints eliminating $A_0$ and $\pi_0$ (with a breaking
of manifest covariance: the reduced phase space contains only $A_i$ and
$\pi_i$). We then quantized the system by introducing Dirac brackets in the reduced phase space. It is appealing to generalize this study for the noncommutative non-Abelian Proca field \cite{Dar}.

\section*{Acknowledgment}

This work has been supported financially by Research Institute
for Astronomy and Astrophysics of Maragha.

\end{document}